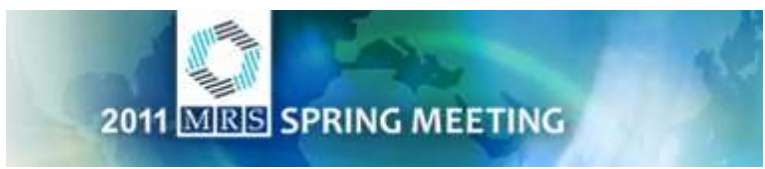

# First Principles Calulations of Defects in Unstable Crystals: Austenitic Iron

SCHOLARONE™
Manuscripts

# First Principles Calculations of Defects in Unstable Crystals: Austenitic Iron

**G.J.Ackland, T.P.C.Klaver and D.J.Hepburn**

**School of Physics, University of Edinburgh, Edinburgh, Scotland EH9 3JZ**

## ABSTRACT

First principles calculations have given a new insight into the energies of point defects in many different materials, information which cannot be readily obtained from experiment. Most such calculation are done at zero Kelvin, with the assumption that finite temperature effects on defect energies and barriers are small. In some materials, however, the stable crystal structre of interest is mechanically unstable at 0K. In such cases, alternate approaches are needed. Here we present results of first principles calculations of austenitic iron using the VASP code. We determine an appropriate reference state for collinear magnetism to be the antiferromagnetic double-layer (AFM-d) which is both stable and lower in energy than other possible models for the low temperature limit of paramagnetic fcc iron. We then consider the energetics of dissolving typical alloying impurities (Ni, Cr) in the materials, and their interaction with point defects typical of the irradiated environment. We show that using standard methods there is a very strong dependence of calculated defect formation energies on the reference state chosen. Furthermore, there is a correlation between local free volume magnetism and energetics. The effect of substitutional Ni and Cr on defect properties is weak, rarely more than tenths of eV, so it is unlikely that small amounts of Ni and Cr will have a significant effect on the radiation damage in austenitic iron at high temperatures.

# INTRODUCTION

First-principles calculations have proved to be a very reliable method of obtaining information about radiation-induced defects in transition metals. In early work on non-magnetic elements such as Mo and V it was shown that the pseudopotential plane wave method reproduced formation energies and barriers to within 0.1eV [1]. It also showed that the strain fields associated with interstitials were much smaller than had been predicted by interatomic potentials, such that reliable values could be calculated with supercells as small as 100 atoms.

Application to steels is a more complicated task on account of their multicomponent nature and the crucial role of carbon in mechanical properties, however work on ferritic Fe and FeCr bcc alloys showed a number of unexpected outcomes. In particular, the binding energy of a Cr atom in Fe is positive, in apparent conflict with the phase diagram which shows a miscibility gap. This conundrum was resolved when it was shown that two Cr atoms in bcc iron strongly repel one another due to antiferromagnetic frustration [2,3], such that a dilute solution of Cr in Fe has positive binding, but a concentrated alloy phase-segregates, because Cr-Cr interactions are unavoidable. This theme of frustration carried over into studies of defects in alloys, however the nonlinear variation of cohesive energy with concentration meant that determining unambiguous energies for quantities such as the binding of an interstitial to a Cr proved impossible in concentrated alloys [4], since the calculated energy had a complex dependence not only on the defect, but also on concentration and the precise location of the atoms.

Austenitic (face-centred cubic) steels have an advantage over their ferritic (body-centred cubic) counterparts in performance and stability at high temperatures. Modelling the high temperature stability of austenitic steel is particularly challenging, because its pure form, gamma-iron is metastable under the zero-temperature conditions typically used in quantum mechanical calculation. The body centred cubic ferromagnetic structure has significantly lower energy than any magnetic arrangement for face-centred cubic (Figure 1). Furthermore, austenite exists naturally only in the paramagnetic state, and paramagnetism is an intrinsically high-temperature phenomenon involving dynamically disordered spins which break local symmetry but produce the fcc structure on average. First principles calculations, use density functional theory to calculate electronic ground states (Born-Oppenheimer dynamics), and so do not include thermal excitations of the magnetic state. Finally, although there remains some debate about whether the paramagnetism is itinerant or involves localized moments on the ions, throughout this paper we will interpret our results through the localized-moment picture.

From a purely numerical point of view there are also difficulties. The Kohn-Sham functional applied in non-magnetic density functional theory has a single minimum with respect to the wavefunctions. By contrast, collinear-magnetic density functional theory, as used for iron, may have up to $2^N$ minima for an N-atom supercell, corresponding to possible permutations of the spin. In practice, most of these will be unstable but one can never be sure that the lowest energy structure has been reached: indeed, since the lowest energy is ferritic, to study austenite one must avoid that structure. The concept of metastability is also slippery, since the numerical tricks used to find the minimum electronic energy do not correspond to physical pathways which the material can follow: the very definition of metastability depends on the computational algorithm.

# EXPERIMENT AND THEORY

## Calculations

The energy of a defect in any material is defined as:

$$E_{def} = E(N, \{M_i\}) - N E_{Fe} - M_i E_{i,ref} \qquad (1)$$

where the first term is the energy of a supercell containing N host atoms and M_i defects of type i. This typically involves a large calculation  The second term is the energy of "austentic" iron and the third term the reference structure for the defect.  For formation energies, this would be zero for topological defects, and the cohesive energy for Ni and Cr impurities.  For migration barrier energies and binding energies, the formation energy is the reference state (substitutional impurity for Ni, Cr, monovacancy or [001] interstitial).

For topological defects (vacancies, interstitials) there is no contribution from the final term. The first challenge of this work is to determine a sensible reference state for austenitic iron.  This is critical, since if the supercell undergoes a transition from one spin state to another, the calculated defect energy will become extensive (i.e. it will scale with the size of the system).

The specific difficulty in austenite is that calculation of defect energies by static relaxation is implicitly a zero-temperature process, and the fcc structure is unstable in iron at zero kelvin.  Furthermore, the ground state of fcc ion is believed to be a spin-spiral structure describable only by non-collinear magnetism and highly sensitive to stresses and strains.  The material of interest is a paramagnetic fcc crystal where the spins are disordered and bear little resemblance to any ordered state.  Thus there is no unambiguously correct way to deal with the magnetism.

In addition to this, there is a general problem of whether to relax the supercell shape.  Again, there are several considerations. If one is considering a periodic array of defects one should relax the cell shape.  If one is considering a random distribution of defects one should relax the volume.  For an isolated defect the lattice parameter is dictated by the material, and so constant volume boundaries are indicated: this is normally the case for radiation-induced defects, although an Eshelby-type elastic correction for finite size[5] of $P^2 V / 2B$ can be applied.  Even relaxation at constant volume leaves open the question of which volume to use: the minimum energy according to the calculation? the extrapolated 0K volume of austenite[6] (a=3.56A)? the volume of the reference state?  A further tricky issue:  should one compare supercells with the same magnetic structure (as is normal for atomic structure, one would not compare defects in bcc with those in fcc) or should the magnetic structure be relaxed to the ground state (as one does in the Born-Oppenheimer approximation for the electronic structure)?

There is no unambiguously correct answer.   There are numerous possible approaches, but the minimal requirement for a reasonable reference state are as follows:

1/ Local minimum  of energy, stable against introduction of defects.
2/ Crystal structure close enough to fcc to be regarded as austenic.
3/ Low energy.
4/ Lattice parameter close to extrapolated values from austenite.
5/ Local moment  comparable with paramagnetic iron.
6/ Stable against single-spin flips
7/ Elastically and dynamically stable

We shall see that the afmI and afmD structures meet these criteria, we take the view that results which are common to both are likely to be reliable.

## Energy and elasticity calculations for candidate reference structures

Calculations were carried out using VASP, a plane wave pseudopotential code, with PAW pseudopotentials for the iron ions with a plane wave cutoff of 400eV and a k-point sampling of $16^3$ for the conventional 4-atom cell, and similar sampling densities for other cells. These settings are converged and appropriate for iron, in particular the PAW is required to reproduce the magnetic moment correctly. Magnetic configurations were established by judicious choice of initial magnetic configuration: we found that there are many minima such that random starting conditions will not find the global magnetics minimum. Full relaxation of the atomic positions was allowed in each case, using a conjugate gradients routine to find the local minimum. For perfect crystal calculations the unit cell was either allowed to relax tetragonally, or constrained to remain cubic. Once the reference structure had been determined, defect calculations were done in fixed supercells. After each calculation the magnetic and atomic structure was re-examined to ensure that there had been no bulk transition to a different phase: for some of the less stable magnetic structures this could happen.

As a preliminary study, the energies of various magnetic phases were determined for a range of densities (see figure 1). The high-spin FM phase used by Jiang and Carter [7] in previous work has the highest energy, and connected via an isostructural phase transition to a lower-spin, lower energy state. It is therefore rather unstable. The AFM-D has the lowest energy, this double-layered (001) structure being more stable than either single layer (AFMI) or triple layer (AFMt).

We also considered the effects of flipping a single spin in a 256-atom unit cell and re-relaxing the structure [8]. Both AFM-1 and AFM-D are stable with one spin flipped, the increase in energy being about 0.05eV [9]. In the high-spin FM state the flipped spin is reduced in magnitude close to zero. In low-spin FM a single spin flip triggers a transformation of the entire spin state to a lower energy configuration.

We also calculated the elastic moduli using the method of Karki et al. [10] (table 1). These moduli are only well defined about the zero stress state which, in the case of AFM-I and AFM-D, is the tetragonally distorted state.

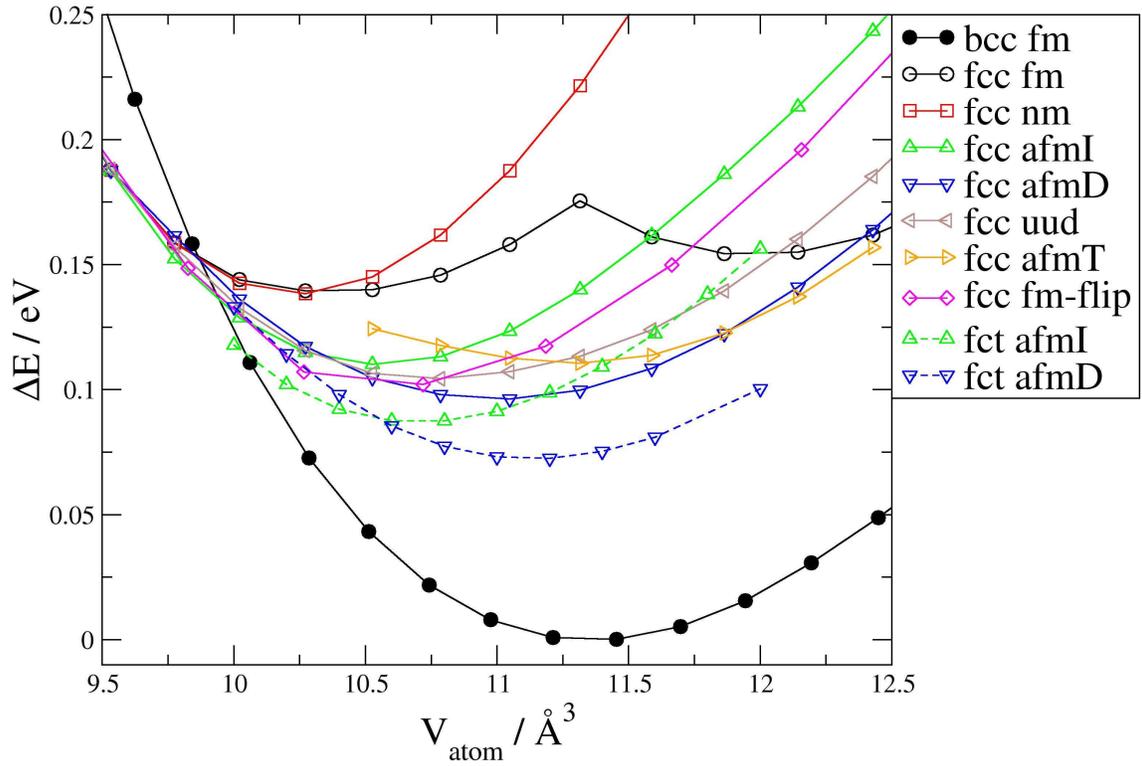

**Figure 1.** Energies of possible reference states for austenite (γ-iron). In addition to those discussed in the text, uud is alternating (001)layers of two spin-up, one spin down, afmT is three up, three down while fm-flip is a conventional fcc cell with one spin reversed..

**Table I.** Elastic constants, relative energies and spins (with tetragonal-distorted values in brackets) for candidate reference structures. Only the fct AFM1 and AFMd structures are mechanically stable at their equilibrium volume. Cubic high spin FM is unstable to an fct distortion, and the fct-distorted cell itself has a negative C' = (C11-C12)/2, revealing its instability to a further orthorhombic distortion.

|  | AFMd (fcc) | FM(fct, high-S) | AFM1 (fct) | AFMd (fct) |
|---|---|---|---|---|
| C11 (GPa) | 224 | 131 | 333 | 212 |
| C33 (GPa) | 119 | 289 | 250 | 210 |
| C12 (GPa) | 147 | 267 | 241 | 211 |
| C13 (GPa) | 81 | 106 | 103 | 92 |
| C44 (GPa) | 87 | 56 | 173 | 73 |
| C66 (GPa) | 108 | 165 | 251 | 203 |
| Energy (meV) | 20 | 31 | 13 | 0 |
| Spin | 1.80 | 2.57 | 1.48 | 199 |

## Comparison of Possible Magnetic Reference States

**Ferromagnetic high-Spin**
In early work for carbon in fcc iron Jiang and Carter [8] took the high-spin ferromagnetic state as a reference. A difficulty with this is that there is also a low-spin state with lower energy and a discontinuous transition between them (Figure 1), and this transition may occur locally without breaking global symmetry. We found this transition (or a transition to a disordered state) could be triggered by topological defects, shear or even by the energy minimization algorithm, and conclude that carbon is a special case, perhaps because it is non-magnetic itself and stabilizes the fcc structure. In defect and impurity calculations we found cases where the impurity induced a transition to the ferrite or to one of the AFM structures. The shear elastic constant $C' = (C_{11}-C_{12})/2$ is negative, indicating that the cubic structure is mechanically unstable. If the implied tetragonal distortion is followed to the fct-FM state, there is still a negative $C'$, in this case leading to an orthorhombic structure.

**Ferromagnetic low-Spin**
The low spin state is mechanically stable and it is possible to run defect calculations with it. However, it is the least similar to the paramagnetic state, both in volume and in moment. For these reasons, we have not used it for the more extensive calculations.

**Non-magnetic**
There is little merit to considering the non-magnetic phase as a model for paramagnetism. The compensation of spins throughout means that there is no magnetic moment on a given atom. As a consequence of this, there is less Pauli repulsion between onsite electrons, and the atoms are "too small". The lattice parameter lies some 10-20% below the spin states.

**AFM-1**
The AFM-I structure has alternating spin on successive (001) planes. This breaks cubic symmetry. This means that the fcc structure is unstable with respect to a tetragonal distortion (it is not even an extremum of the energy). It is possible to relax to the tetragonal minimum, which is elastically stable. The density is significantly higher than the paramagnetic phase. Furthermore, several possible defects we which examined induced a change of spin-state from cubic AFM-1. We note that although the energy of AFM-1 is higher than for AFM-D for pure iron, in some concentrated alloys the situation is reversed.

**AFM-D**
The AFM-D structure has alternating spin on paired double layers. It also breaks cubic symmetry, such that one has to use tetragonal supercells. It has the advantage of being the lowest energy collinear structure we have found, and a lattice parameter and spin much closer to the paramagnetic state. Its equilibrium volume is closer to the 0K extrapolated austenite. However, it has a very low $C'$ shear modulus, which becomes negative at increased volume. For calculations at negative pressure (e.g. the density of gamma-iron under temperature conditions where it is actually stable) the ground state of AFM-D has a further, orthorhombic distortion. For $Fe_{70}Cr_{20}Ni_{10}$ composition it proved impossible to stabilize the AFM-D ordering at all.

**Randomized spins**
It would be possible to randomize the spins on each site, or to consider a set of "special quasi-random structures" to represent all possible paramagnetic arrangements in the same way as one deals with random alloys. However, preliminary calculations showed that defect calculations were extremely sensitive to spin-flips: details of the minimization process could change the magnetic state leading to differences in energy of tenths of eV for apparently identical configurations. Similarly, the energy of the reference state is not well defined: an average over SQS structure is required, and since each SQS structure breaks symmetry in a different way, some decision must be reached regarding the relaxation of cell and atoms, bearing in mind the possibility of collapse to ferritic states. Thus this approach was found to be impractical.

**Non-collinear magnetism**
It is possible to use non-collinear magnetism and start by setting up a spin-spiral commensurate with the cell, or set up randomized initial spins. We calculated energies for a set of representative configurations: vacancy, Cr and Ni substitutionals, <001> and octahedral interstitials, a Cr-Ni mixed-interstitial dumbbell, and Ni-vacancy binding. After relaxation from a random spin-state, only the last of these retained non-collinearity. Even in this case, the effect on the binding energy was marginal, certainly less that other sources of error. We therefore concluded that including non-collinear effects is unnecessary in this case.

## Defect Calculations

We have extended our calculations of austenitic iron to encompass point defects, self-interstitials, substitutional Cr and Ni, these being commonly-used components of steels. Given the instability of the FM states against single spin flips, we concentrate on AFM-1 and the two AFM-D states with supercells based on the minimum energy cubic or tetragonal cell as the best reference state for impurities in gamma-iron. In each case the reference state is the isolated substitutional atom of Cr or Ni in Fe. Because of the symmetry-breaking effect of the AFM ordering, there are more distinct defects and migration barriers than in a homogeneous system. We have calculated many of them: some are unstable to reconstructions, and only the lowest ones are considered.

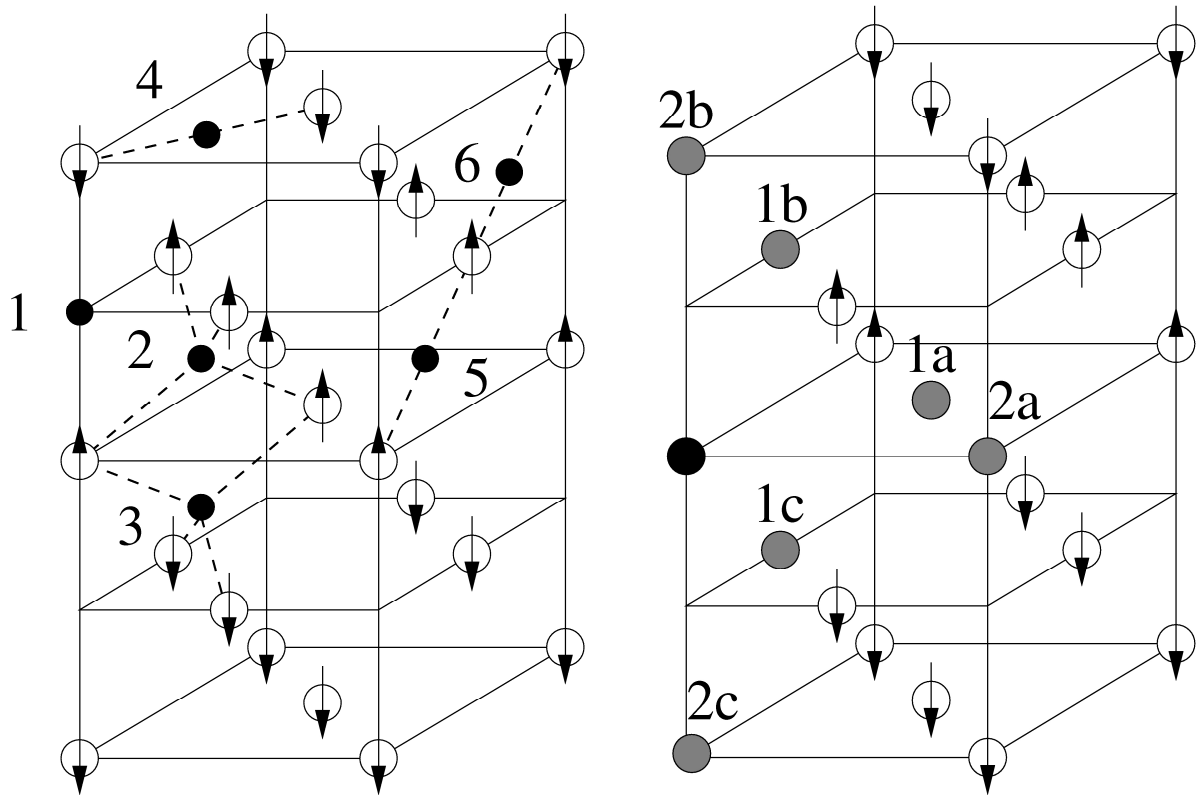

**Figure 2.** *Numbering scheme for defect positions considered in this work: Left, interstitial sites; right, neighbour positions from defect (unmarked black circle). Arrows show relative spin direction for collinear AFMd reference state.*

## Point Defects

The various defects considered here are shown in Figure 2, and the defect energies given in Table 2. The most striking feature is the difference between the different reference states. Even the vacancy formation energy differs by 0.15 eV. The migration barrier is lowest for those configurations which do not require a spin flip, but varies by up to 0.8eV. Vacancies on adjacent sites are bound, but again the strength of binding depends on the spin state. Turning to the interstitial, the [001] dumbbell is the most stable structure, with various possible migration paths lower than those available to the vacancy. Again, there is a strong dependence on the reference state, with cubic giving lower formation energies: a consistent difference of some 0.3eV, significant, albeit much lower than the strain energy. The interstitial atom breaks the fcc symmetry and allows the unstable fcc-AFMd to relax towards the stable fct-AFMd. This is evidence in favour of using the lowest energy state (AFMd).

The position relative to the spin layers also has an effect . The magnetization around the defects increases with the free volume around the atom. Thus atoms around the vacancy, and in the tensile region beside the interstitial, have enhanced magnetization. Meanwhile those in the compressive region have reduced magnetization. The AFMd and AFM1 structures are stable around these defects. Of course, the interstitial atoms themselves acquire a magnetization, and in the case of the [001] dumbbell the magnetic defect extends to the first neighbour shell, flipping four spins so as to have four (enhanced) parallel spins in the tensile plane, with adjacent compressive layers being antiparallel.

## Effect of Cr and Ni Impurities

We define the solution energy for adding a Cr or Ni atom to Fe relative to the pure fcc Ni or Cr. This eliminates systematic energies arising from the poor description of the free atom by non-magnetic DFT, which is the default reference state in VASP. We also calculated interactions between these substitutional impurities and with point defects. Results are shown in table 3. The main feature of this table is the small value of all the energies, rather close to the difference between the two reference states. It appears that Ni and Cr are almost invisible to one another. The only process in which they behave significantly differently from the host is in the exchange with a vacancy, where the Cr has a lower barrier than for pure Fe, while the Ni a higher one. Even this result is tempered by the very high energy to exchange Fe between layers of opposite spin, although this is probably anomalous, since in collinear magnetism the symmetry requires that the migrating atom has zero spin at the symmetry point, a high energy state for iron. When this constraint is relaxed so that the atom retains a moment the energy barrier is lowered by about 0.5eV (in AFM1), assuming that there is no energy barrier to flipping the spin.

**Table II.** *Energies of point defects calculated with AFM-D and cubic or tetragonal supercells, fixed volume with all atomic positions and magnetic moments relaxed. $E_f$, $E_m$ and $E_b$ refer to formation, migration and binding energies respectively. V refers to vacancies. Numbers in brackets in the first column refer to defect positions given in figure 2. rlx() indicates that the defect relaxes to the position shown in the bracket. The corrections for using constant volume cell, calculated from linear elasticity theory, are negligible for all cases apart from the interstitials, where they are of order 0.05eV, meaning the choice of constant volume vs constant pressure relaxation introduces an uncertainty of around 2%*

| Defect | AFMd (c) | AFMd (t) | AFM1 (t) |
|---|---|---|---|
| E_f,V | 1.672 | 1.819 | 1.953 |
| E_m,V (1a) | 1.046 | 0.743 | 0.622 |
| E_m,V (1b) | 0.712 | 1.098 | |
| E_m,V (1c) | 1.268 | 1.581 | 1.724 |
| E_b,VV (1a) | 0.205 | 0.037 | 0.063 |
| E_b,VV (1b) | 0.056 | 0.127 | |
| E_b,VV (1c) | 0.075 | 0.175 | 0.046 |
| E_b,VV (2a) | 0.022 | -0.064 | 0.073 |
| E_b,VV (2b) | -0.079 | -0.018 | -0.090 |
| octa (1) | rlx (8) | Rlx(8) | 4.353 |
| Tetra uu (2) | 3.581 | 3.864 | |
| Tetra ud (3) | 3.332 | 3.663 | 4.322 |
| [110] crowdion (4) | rlx (3) | rlx(3) | 4.799 |
| [011] crowdion (5) | 3.771 | 4.255 | |
| [01-1] crowdion (6) | 3.874 | 4.168 | 4.818 |
| [100] dumbbell (7) | 2.978 | 3.316 | 3.531 |
| [001] dumbbell (8) | 2.790 | 3.195 | 3.615 |
| [110] dumbbell (9) | 4.289 | 4.311 | 4.803 |
| [011] dumbbell (10) | rlx (8) | Rlx(8) | Rlx |
| [111] dumbbell (11) | rlx (3) | Rlx(3) | 4.559 |

**Table III** *Binding energies of Ni, and Cr to vacancies and interstitial clusters. Interstitial configurations other than [001]-type are higher in energy. Note that the higher symmetry in AFM1 means that some configurations do not exist*

| | AFMd (c) | AFMd (t) | AFM1 (t) |
|---|---|---|---|
| Substitutional energy Ni | -0.033 | 0.084 | 0.167 |
| Substitutional energy Cr | 0.106 | 0.268 | 0.047 |
| V-Ni  (1a) | 0.043 | 0.056 | 0.089 |
| (1b) | 0.013 | 0.027 | |
| (1c) | -0.036 | 0.016 | 0.042 |
| (2a) | 0.026 | -0.002 | 0.011 |
| (2b) | -0.048 | -0.011 | -0.010 |
| (2c) | -0.051 | -0.005 | |
| V-Cr (1a) | -0.087 | 0.004 | 0.030 |
| (1b) | -0.066 | -0.075 | |
| (1c) | -0.088 | -0.091 | -0.079 |
| (2a) | -0.025 | -0.016 | -0.052 |
| (2b) | -0.039 | -0.066 | -0.075 |
| (2c) | -0.036 | -0.004 | |

| Ni-Ni  Ni-Cr  Cr-Cr  binding of substitutional impurities | | | |
|---|---|---|---|
| 1a | 0.024, -0.016, -0.044 | 0.055, 0.025, -0.027 | 0.107,0.050,-0.061 |
| 1b | -0.004, -0.018, -0.005 | 0.024, -0.012,-0.012 | |
| 1c | -0.043, -0.021,  -0.093 | -0.014, 0.027, -0.098 | 0.031,-0.004,-0.071 |
| 2a | 0.036 , -0.021 0.043 | 0.017, -0.016, 0.023 | 0.067,0.022,0.008 |
| 2b | -0.045, -0.021 -0.001 | -0.014, -0.015,-0.011 | 0.055,-0.025-0.009 |
| Vacancy  migration (Ni, Fe, Cr) | | | |
| 1a | 1.304, 1.046, 0.846 | 0.891, 0.743 , 0.560 | |
| 1b | 0.883, 0.711, 0.572 | 1.172, 1.048,0.742 | |
| 1c | 0.932, 1.268, 0.657 | 1.179, 1.581, 0.844 | |
| Mixed dumbbell Interstitials | | | |
| 100 Ni, Fe, Cr | 3.416, 2.977, 3.049 | 3.717, 3.315, 3.385 | 4.112,3.531,3.583 |
| 001 uu Ni, Fe, Cr | 3.069, 2.790, 2.933 | 3.229, 3.195, 3.197 | |
| 001 ud Ni, Fe, Cr | 3.097, 2.790 2.850 | 3.469, 3.195,  3.270 | 4.116,3.615,3.267 |

## DISCUSSION AND CONCLUSIONS

We have calculated defect energies for a number of configurations using three possible representations for AFM austenite. The choice of relaxation strategy is clearly important: our results suggest that the tetragonally distorted AFM structures present the best reference structure to use for point defects in pure gamma-iron. Failure to pick a minimum energy supercell can lead to spurious "energy gains" when the unstable structure relaxes around a defect. In preliminary runs this problem was so pronounced that most reference structures were rejected altogether. In the result reported here a mild form of the problem is illustrated best by the comparison of interstitials in AFMd (c) with AFMd (t). Although the numbers for AFMd (c) are plausible, the fact that defects have consistently lower energies can only be attributed to accompanying relaxation of the structure.

The large differences between states which would be identical in fcc need to be interpreted: could such configurations exist locally in high temperature austenite, such that the formation and migration energies should be taken as the minimum of those we have found? Or should we consider these as extreme cases in a continuum of complicated local spin states, and take a mean? Whether we are really modelling austenite remains open to question.

Comparing AFMd and AFM1 gives us a good indication of the uncertainty introduced by the choice of magnetic state. This is also shown by the differences between configurations which would be identical in fcc, but have their symmetry broken by the magnetic ordering. There are several conclusions which are robust despite these approximations. The solution energy of Cr and Ni in fcc iron is small, and their binding to each other is weak, implying that local ordering has little role to play. Interstitials are repelled from Ni and vacancies from Cr, but in the dilute alloy this will have little effect. Cr binds with a strength of a few hundredths of an eV to interstitials and nickel binds similarly to vacancies. In a dilute alloy this may be enough to affect the evolution of primary radiation damage. In a more concentrated alloy most vacancy sites will be adjacent to at least one Ni among its 12 neighbours, and this will limit the pinning effect. For interstitials the chromium may prove quite a strong obstacle, since its favoured site is on the dumbbell itself.

These calculations provide energy data which should be of use to initiate multiscale modelling, either in the making of interatomic potentials for steels [12-15] or for monte carlo simulations [16-17].

## ACKNOWLEDGEMENTS

This work was sponsored by the EU under the FP7 PERFORM-60 and GETMAT programs. Some of the results were presented previously in the the Second International Workshop on Structural Materials for Innovative Nuclear Systems (SMINS-2) in Daejeon. We also thank the EPSRC for computer time under the UKCP programme GR/S14658.